# Data as an economic good, data as a commons, and data governance

Nadya Purtova and Gijs van Maanen

*Utrecht University School of Law, Utrecht University, Utrecht, the Netherlands*

Newtonlaan 201 (4A), 3584 BH Utrecht

n.n.purtova@uu.nl

# Data as an economic good, data as a commons, and data governance

*[accepted for publication in* Law, Innovation, and Technology *16(1)]*

This paper provides a systematic and critical review of the economics literature on data as an economic good and draws lessons for data governance. We conclude that focusing on data as an economic good in governance efforts is hardwired to only result in more data production and cannot deliver other societal goals contrary to what is often claimed in the literature and policy. Data governance is often a red herring which distracts from other digital problems. The governance of digital society cannot rely exclusively on data-centric economic models. We review the literatures and the underlying empirical and political claims concerning data commons. While commons thinking is useful to frame digital problems in terms of ecologies, it has important limitations. We propose a political-ecological approach to governing the digital society, defined by ecological thinking about governance problems and the awareness of the political nature of framing the problems and mapping their ecological makeup.

## 1  Introduction

Data governance has been on top of the European regulatory agenda for a decade, revived with the 2012 data protection reform and culminating in the avalanche of other regulatory instruments proposed and adopted in the past two years.[1] Much of the data governance

---

[1] This is an incomplete list of the proposed and adopted EU legislation with relevance for data governance: Regulation (EU) 2016/679 of the European Parliament and of the Council of 27 April 2016 on the protection of natural persons with regard to the processing of personal data and on the free movement of such data, and repealing Directive 95/46/EC (General Data Protection Regulation) *OJ L 119, 4.5.2016,* p. 1–88; Directive (EU) 2019/944 of the European Parliament and of the Council of 5 June 2019 on common rules for the internal market for electricity and amending Directive 2012/27/EU *OJ L 158, 14.6.2019,* p. 125–199*;* Directive (EU) 2019/1024 of the European Parliament and of the Council of 20 June 2019 on open data and the re-use of public sector information *OJ L 172, 26.6.2019, p. 56–83;* Regulation (EU) 2022/868 of the European Parliament and of the Council of 30 May 2022 on European data governance and amending Regulation (EU) 2018/1724 (Data Governance Act) *OJ L 152, 3.*6.2022, p. 1–44 entered into force on 23 June 2022 and will be applicable from September 2023; Regulation (EU) 2022/2065 of the European Parliament and of the Council of 19 October 2022 on a Single Market For Digital Services and amending Directive 2000/31/EC (Digital Services Act) OJ L 277, 27.10.2022, p. 1–102, effective from January 2024; Regulation (EU) 2022/1925 of the European Parliament and of the Council of 14 September 2022 on contestable and fair markets in the digital sector and amending Directives (EU) 2019/1937 and (EU) 2020/1828 (Digital Markets Act) *OJ L 265,* 12.10.2022, p. 1–66, in effect from June 2023; at the time of writing, Council has adopted its general approach on the Artificial Intelligence Act Data Act

literature[2] informing those and future initiatives builds analyses of governance options such as data access and sharing based on assertions about the nature of data as an economic good: data is either characterised as a club good,[3] or more recently as a common-pool resource,[4] infrastructure,[5] labour[6] or capital.[7] The literature on governing data as the commons and the related thinking about collective forms of data governance have gained huge momentum in recent years.[8] Economic models associated with those classifications are often adopted as recipes for successful data governance, where the success is measured by the effects of the proposed governance strategies on the growth of the digital economy and innovation, health, energy sustainability, empowerment of individuals as consumers, patients, or data subjects,[9]

---

(Proposal for a Regulation of the European Parliament and of the Council laying down harmonised rules on artificial intelligence (Artificial Intelligence Act) and amending certain Union legislative acts - General approach, adopted on 25 November 2022); Proposal for a Regulation on the European Parliament and of the Council on harmonised rules on fair access to and use of data (Data Act) COM/2022/68 final; Proposal for a Regulation of the European Parliament and of the Council on the European Health Data Space (COM/2022/197 final).

[2] We understand "data governance" broadly as a multidisciplinary body of literature addressing various aspects of how data should be dealt with in relation to societal goals, from innovation to ensuring production in the data-driven economy to the protection of privacy.

[3] Literature discussed in Section 4.

[4] Literature discussed in Section 5.

[5] J. M. Nolin, 'Data as oil, infrastructure or asset? Three metaphors of data as economic value' (2020) 18 Journal of Information, Communication and Ethics in Society 28 <https://doi.org/10.1108/JICES-04-2019-0044> accessed 6 April 2023. Charlotte Ducuing, 'Data as Infrastructure? A Study of Data Sharing Legal Regimes' (2020) 21 Competition and Regulation in Network Industries 124 <https://doi.org/10.1177/1783591719895390> accessed 6 April 2023; Tommaso Fia, 'An Alternative to Data Ownership: Managing Access to Non-Personal Data through the Commons' (2021) 21 Global Jurist 21 181 https://doi.org/10.1515/gj-2020-0034 accessed 6 April 2023.

[6] E.g. Jérôme Denis and Samuel Goëta S, 'Rawification and the Careful Generation of Open Government Data' (2017) 47 Social Studies of Science 604 <https://doi.org/10.1177/0306312717712473> accessed 6 April 2023; Erna Ruijer and others, 'Open Data Work: Understanding Open Data Usage from a Practice Lens' (2018) 0 International Review of Administrative Sciences 1.

[7] Alex Pentland, 'Building the New Economy. What we need and how to get there' in Alex Pentland, Alexander Lipton and Thomas Hardjono (eds), *Building the New Economy: Data as Capital* (The MIT Press, 2021); Jathan Sadowski, 'When Data Is Capital: Datafication, Accumulation, and Extraction' (2019) 6 Big Data & Society <https://doi.org/10.1177/2053951718820549> accessed 6 April 2023.

[8] Literature in 4.2, but also Joan López, Aaron Martin, Linnet Taylor and Siddharth Peter De Souza, 'Governing data and artificial intelligence for all: Models for sustainable and just data governance', study written for the Panel for the Future of Science and Technology, European Parliamentary Research Service, July 2022, available at <https://globaldatajustice.org/gdj/category/publications/> last accessed 9 December 2022. The study advocates for shifting the data governance focus towards "collective will and decision-making on the part of societal groups, combined with the normative orientation toward public value" (65).

[9] e.g. Communication from the Commission to the European Parliament, the Council, the European Economic and Social Committee and the Committee of the Regions "A European strategy for data" COM(2020) 66 final, Brussels, 19.2.2020, 2 et seq.

and other broad societal goals. What this academic and policy literature is often missing is a more nuanced view of the "data as an economic good" way of thinking where the advantages of using the economic classifications to achieve regulatory objectives are balanced with the awareness of the limitations of those classifications. This article addresses this blind spot in the data governance literature.

This contribution does three things. First, it offers a systematisation of the broad range of scholarship on data as an economic good. The academic discourse on data as an economic good and how it should be governed is ridden with complexity and terminological confusion, which stands in the way of its productive use in academic, policy and practical contexts. For instance, the dominant economic stance on the nature of information goods is that information is a public good.[10] A more recent take is that data as a sub-type of an information good is a club good.[11] Yet, others argue that data is a common-pool resource or a commons.[12] In the commons scholarship, some construe data as a common-pool resource based on its inherent characteristics (its subtractability and difficulty to exclude beneficiaries),[13] while others argue that the collective data management or political claims on data that communities have, make it a commons.[14] These are mutually exclusive statements and leave a reader at a loss about what argument to follow. This paper offers a systematic way to look at this perplexity and make sense of it, which will allow our readership to use the reviewed literature more productively.

Second, we interrogate to what extent, and for which purposes the reviewed strands of literature are helpful or unproductive to guide governance efforts. Our primary criticism is that

---

[10] Literature in 4.
[11] Literature in 4.
[12] Literature in 5.
[13] E.g. Priscilla M. Regan, 'Privacy and the Common Good: Revisited' in Beate Roessler and Dorota Mokrosinska (eds), *Social Dimensions of Privacy: Interdisciplinary Perspectives* (Cambridge University Press 2015), 392 et seq.; J.J. Zygmuntowski, L. Zoboli and P.F. Nemitz, 'Embedding European Values in Data Governance: A Case for Public Data Commons' (2021) 10 Internet Policy Review <https://policyreview.info/articles/analysis/embedding-european-values-data-governance-case-public-data-commons> accessed 12 October 2021.
[14] Literature reviewed in 6.1.

- contrary to what some suggest - adopting data as an economic good as a focus of the governance efforts is hardwired to only produce governance strategies that will facilitate the provision of more or better-quality data. This is in line with a broader criticism of the performativity of economic analyses, which do not merely study but also shape the world as an economy.[15] If the regulatory objective is to attain other societal goals beyond data provision, e.g. to protect privacy and other fundamental rights and interests relating to data, empower individuals, or even strengthen the digital economy, the focus on data as an economic good is not productive.[16]

Finally, we reflect on what our analysis means and draw lessons for and about data governance. We use the broader term "governance" rather than "regulation" to reflect the focus of the paper beyond but with relevance for state regulation, and the fact that some of the literatures we review advocate for grass-roots collective governance rather than state regulation.[17]

We draw five lessons for and about data governance and governing the digital society in general.

(1) We submit that data can simultaneously be a club good and a part of a larger common-pool resource. We call this a "dynamic classification of data" because one can switch between the options depending on the purposes of one's analysis. The club good characterisation guides how to ensure that enough data of sufficient quality is available. The commons analytical framework is a better match where something else needs to be provided and where data is instrumental to that something else's sustainability.

---

[15] e.g. Timothy Mitchell, 'The work of economics: how a discipline makes its world' (2005) 46 European Journal of Sociology / Archives Européennes de Sociologie 297 <https://doi.org/10.1017/S000397560500010X> accessed 6 April 2023.

[16] See 4 and 7.2.

[17] We understand governance as defined by five propositions: "(1) Governance refers to a complex set of institutions and actors that are drawn from but also beyond government. (2) Governance recognises the blurring of boundaries and responsibilities for tackling social and economic issues. (3) Governance identifies the power dependence involved in the relationships between institutions involved in collective action. (4) Governance is about autonomous self-governing networks of actors. (5) Governance recognises the capacity to get things done which does not depend on the power of government to command or use its authority." Gerry Stoker, 'Governance as theory: five propositions' (1998) 50 International Social Science Journal, 17. But compare this to the definition of governance as "a government's ability to make and enforce rules" in F. Fukuyama, 'What is governance?' (2013) 26 Governance 347, 350 <https://doi.org/10.1111/gove.12035>.

(2) It is not productive to focus on the governance of *data* as an economic good if the objective is to attain a societal goal other than the provision of data of sufficient quantity and quality. The reviewed neo-classical economic models are hardwired to ensure the provision of goods (i.e. that there is enough and sufficient quality of those available) and hence can only ensure the provision of data but not other broader societal goals, such as innovation, digital economy, privacy, and others, even when those are associated with data. In this sense, governance of the digital society should not be (exclusively) data governance. Data governance can become a red herring which distracts from other digital problems.

(3) Neo-classical economics engagement with data as an economic good detracts from the complexity of data relations, and the role data plays for collectives.

(4) The grass-roots collective (data) governance initiatives should not become bandaids to problems where public regulation fails or is non-existent. While collective governance and citizen self-organisation have an important place in the governance toolbox, they should not take on the burden better placed on the state regulation. This is especially the case when the protection of fundamental rights and issues transcending the boundaries of small communities are concerned.

(5) We propose a political-ecological approach to governing the digital society, which is defined by ecological thinking about governance problems and the awareness of the political nature of framing the problems and mapping their ecological makeup.

Economics operates with many classifications of economic goods. There is a classification into experience, credence and search goods based on how easy or difficult it is for a consumer of the good to assess its quality either prior to or after consumption. Our review includes only the literature on the classification along the axes of rivalry and excludability. This choice is made for two reasons. First, as will become apparent from the analysis below, this classification is frequently connected to the governance options in the data governance literature. Second, reviewing all classifications is simply impossible in one paper. We also do not consider literature conceptualising data as a particular kind of a good, e.g. infrastructure, labour or capital. The consequence of this choice is that our critique of economic thinking about data is only limited to the classifications along the axes of subtractability and excludability and cannot be generalised to apply to other strands of economic thinking.

## 2    Agreeing on terms: good, data, information, and knowledge

We begin by introducing some key terms of significance for the analysis: good, data, information, and knowledge. We use the term "good" the way it is understood in economics, where it has a very broad meaning. A "good" in economics stands for "all desirable things, or things that satisfy human wants," both material and immaterial.[18] This is different from the colloquial meaning of the word "good" as an object of sale or another market exchange. Despite the market connotations that the word "good" might have and despite the very real nature of data markets, the focus of this paper on data as an economic good does not imply that we advocate for data market or commodification.

Another key term is "data." Modern understanding of data is closely intertwined with the computer, information technology, and information theory.[19] In the digital context, which

---

[18] Alfred Marshall, *Principles of Economics: Unabridged Eights Edition* (Cosimo Classics 2009), 45.
[19] D. Rosenberg 'Data before the Fact' in L. Gitelman (ed) *Raw Data Is an Oxymoron* (The MIT Press, 2013), 34.

is increasingly a default setting of discourses on data governance, data is understood as a digital representation of information[20]: a result of quantification of information,[21] or "information in numerical form."[22] ,[23] in European regulation,[24] and in this paper. This is the definition primarily adopted by the data governance and economics literature

There exist many classifications and categorizations of data: metadata, raw vs processed data, synthetic vs "real" data, personal vs non-personal data, etc. Our analysis is generally agnostic of these distinctions because they are of no significance to the economic characterisation of data as a good along the axes of excludability and subtractability discussed further in this paper.[25]

The notions of data and information are closely linked, as data is defined through information. It is not easy to provide one definition of information. An entire branch of economics called "information economics" and, in particular, game theory addresses how information and different information structures shape the interaction of economic actors.[26] More general economic theory, which this paper reviews in the following sections, considers information as an economic good. Yet, economics does not have its own definition of information. This void is filled by many other disciplines as diverse as philosophy, psychology, biology, cybernetics, and many others that offer their own definitions of information, resulting in conceptual chaos or what Burgin calls 'information studies perplexity'.[27] While the existing definitions of information are dozens, they can be roughly classified into three approaches: semantic, syntactic, and functional.[28]

---

[20] But see critical data studies literature where the performative role of data is emphasised, e.g. C. D'Ignazio and L.F. Klein, *Data Feminism* (The MIT Press 2020).

[21] Raphael Gellert, 'Comparing definitions of data and information in data protection law and machine learning: A useful way forward to meaningfully regulate algorithms?' (2020) Regulation & Governance <https://onlinelibrary.wiley.com/doi/epdf/10.1111/rego.12349> accessed 6 April 2023.

[22] Rosenberg (n19) 33.

[23] E.g. "[d]ata is information that can be encoded as a binary sequence of zeroes and ones." In Maryam Farboodi and Laura Veldkamp, 'A growth model of the data economy' (2021) NBER Working paper Series, Working Paper 28427 <http://www.nber.org/papers/w28427> last accessed 4 December 2021, 2.

[24] Arts 2(1) of the Data Act, Data Governance Act, DMA and European Health Data Spaces Regulation define data as "any digital representation of acts, facts or information and any compilation of such acts, facts or information, including in the form of sound, visual or audio-visual recording".

[25] Section 3. Some of these distinctions are problematic. See, Lisa Gitelman (eds) *Raw data is an oxymoron* (MIT Press 2013) on why there is no such a thing as raw data; N. Purtova, 'The Law of Everything. Broad Concept of Personal Data and Future of EU Data Protection Law' (2018) 10 Law, Innovation and Technology 40, on the problems with the distinction between personal and non-personal data. It is also debatable what makes data "real". Under some circumstances, even synthetic data can become very real in its effects (e.g. it can be used in automated decision-making and will be personal data if used with a purpose or effect to assess or influence a natural person who is identified or identifiable).

[26] Eric Rasmusen, *Games and Information: An Introduction to Game Theory* (Wiley-Blackwell 3rd ed. 2001).

Within the semantic approach, information is contingent on the presence of meaning. The meaning of information is relative and depends on the receiver of information. Hence, information is contingent on the presence of human cognition. When the meaning is understood and internalised, this results in knowledge. Within this school of thought, several analyses have adopted a General Definition of Information ('GDI'): *'information is data + meaning'*.[29]

Within the syntactic approach, whether information has meaning is irrelevant. A notable example of syntactic understanding of information is Shannon's mathematical theory of information formulated in the context of signal transmission. It defines information as "the statistical probability of a sign or signal being selected from a given set of signs".[30] Another example of the syntactic approach to information is the so-called naturalisation of information in the works of physicists, engineers and biologists, where information is seen as belonging to the natural world, expressed in the specific biological structures,[31] the building material of the universe[32] and not anthropocentric.[33]

---

[27] M. Burgin, *Theory of information. Fundamentality, diversity and unification* (World Scientific Publishing, 2010) 6.

[28] It exceeds the scope of this paper to provide a comprehensive review of the information literature and all possible definitions of information. For a comprehensive cross-disciplinary overview of the various meanings of the term 'information' see Rafael Capurro and Birger Hjørland, 'The concept of information' in *Theorizing information and information use*, 343-411 and M. Burgin, *Theory of information. Fundamentality, diversity and unification* (World Scientific Publishing, 2010). For an alternative brief overview of the various approaches to information see Gellert (n18).

[29] L. Floridi, 'Is Information Meaningful Data?' (2005) 70 Philosophy and Phenomenological Research 351.

[30] F. Machlup, F. 'Semantic quirks in studies of information', in F. Machlup & U. Mansfield (eds), *The study of information: Interdisciplinary messages* New York: Wiley 1983), 658.

The third approach, which we call "functional", reconciles syntactic and semantic approaches and defines information not by what it is but what it does. Information affects reality by either changing the knowledge about it, constituting or changing it. For instance, according to Floridi, information refers to three mutually compatible phenomena: information *about* reality (semantic information); information *as* reality (e.g. "as patterns of physical signals" such as DNA, or fingerprints); and information *for* reality ("instructions like genetic information, algorithms, orders or recipes").[34] Capurro and Hjørland argue that the syntactic and semantic definitions of information are not mutually exclusive but have a common core, i.e. selection: "we state a resemblance between interpreting meaning and selecting signals. The concept of information makes this resemblance possible."[35] Therefore the various definitions of information should be considered in a relationship of Wittgensteinian "family resemblance"[36] and not as correct or incorrect, but rather more or less useful in a certain context.[37]

We use "information" in the latter, functional, meaning which lends itself well for analysis of the economics literature: on the one hand game theory relies game theory on the semantic understanding of information as it concerns itself with the decision makers that *know* that their decisions affect each other and engage in strategic reasoning on the basis of information available to them. On the other hand, the general economic theory concerning the classification of goods as rival and excludable is agnostic of whether information has meaning. Therefore this analysis is based on the functional understanding of information.

---

[31] B. O. Küppers, 'The context-dependence of biological information', in K. Kornwachs & K. Jacoby (eds), *Information. New questions to a multidisciplinary concept* (Akademie Verlag 1996), 140, quoted in Capurro (n28).
[32] Burgin (n27), and in-text references.
[33] Capurro (n28), 358 and 360 et seq.
[34] Luciano Floridi, *The Philosophy of Information* (Oxford University Press 2011) 30.
[35] Capurro (n28) 368.
[36] Ibid.
[37] Ibid. 358.

Knowledge is an information-related concept. Knowledge is "the accumulation and integration of information received and processed by a recipient."[38] It is consistent with the semantic and functional approaches to defining information through the changes it brings - when processed by human cognition - in knowledge.[39] The so-called information- or knowledge hierarchy developed by the knowledge management scholars such as Zeleny[40] and Ackoff[41] lay connections between data, information and knowledge. The hierarchy defines "data as 'know-nothing', information as 'know-what', knowledge as 'know-how', and [...] wisdom as 'know-why'".[42] While this contribution is focused on data, the concept of knowledge will become relevant in discussions about the role of data in "the scientific knowledge commons".[43]

## 3    Economic classification of goods: a brief introduction

A major way of classifying goods in neo-classical economics has been along two axes: 1) rivalry of consumption of a good and 2) the possibility to exclude others from a particular good or enjoying its benefits. A good is rival if consumption by one individual prevents others from consuming the same good at the same time or at all. A good is "excludable" if its consumption can be prevented, e.g. because of not having paid for it. Historically speaking, economists distinguished two types of goods which were diametrically opposed in both of these dimensions: pure private and pure public goods.[44] Pure private goods, mostly tangible goods, are rival as well as excludable. If one eats an apple, that apple is not available to anyone else, and it is easy to exclude people from the possibility of consuming an apple, e.g. by making it physically not accessible. The fact that consumption of such a good is rival as well as the ability of one to exclude others from consumption implies that they can be enjoyed by one person alone without interference by others or enjoyment spillovers to others. Hence, their utility cannot be shared. However, the good (and the benefits that come along with it) could be fully or partially transferred, e.g. sold, to someone else who would then be enjoying the (share of the) private good.

---

[38] Burgin (n27) 192.
[39] Ibid. 331 et seq. and the literature discussed there.
[40] M. Zeleny, 'Management support systems: towards integrated knowledge. Management' (1987) 7 Human Systems Management 59.
[41] R. L.. Ackoff, 'From Data to Wisdom' (1989) 16 Journal of Applied Systems Analysis 3.
[42] Burgin (n27) 186, citing M. Zeleny, 'Management support systems: towards integrated knowledge management' (1987) 7 Human Systems Management 59. According to some accounts, this hierarchy is problematic, since it implies that data is deprived of meaning and information is defined by meaning (e.g. Rosenberg (n19) 33, and the reviewed literature on syntactic and functional approaches to information).
[43] Sections 5.2 and 5.3.

Pure public goods are non-rival and non-excludable. National defence is often used as an example of a pure public good: the fact that one individual enjoys the benefits of national defence does not make any less of it available to others, and it is prohibitively costly to prevent country residents from using national defence benefits. Despite its name, a public good can be provided both by the public or private sector, or simply already exist independently of any action by anyone (e.g. sunlight). However, when a public good needs to be actively (and costly) provided, a free rider problem arises. Because no one can be excluded from enjoying the benefits of a public good, also those that do not contribute to the provision of the public good (the "free riders") benefit from it. There are no strong incentives for any individual to contribute to the provision of the public good leading to its under-provision. To solve the free rider problem many public goods are provided by the public sector as the state can exercise coercive power for contributions via taxation and fund the public good.

With time, however, empirical research demonstrated that some resources do not easily fit within this binary – public vs private – classification. Many public goods, for instance, only exhibit the characteristic of non-rivalry and non-excludability to a certain extent, such as within a certain group or space (e.g. national defence in a particular country), or for a limited time (the music played by a street musician is non-rival and hence a public good, but this would not continue to be the same if everyone stops to listen at the same time and stays around). Thus, public goods may only fulfil the two conditions on some level, rendering them so-called "local public goods".

---

[44] R. Cornes and T. Sandler, *The Theory of Externalities, Public Goods, and Club Goods* (Cambridge University 2nd ed, 1996), 5 et seq.

To reflect this, some changes were made in the neo-classical classification of goods.[45] "Rivalry of consumption" was replaced with "subtractability of use", and subtractability and excludability were operationalized as varying from low to high as opposed to binary, 1 or 0, characterizations. Goods displaying characteristics of both public and private goods were initially considered impure public goods but eventually were distinguished into two new categories: club (or toll) goods and common-pool resources. The resulting economic classification of goods is presented in Figure 1.[46]

On a global scale only a few things, such as world peace, may truly fulfil the two criteria of a public good and could therefore be called a "global public good". Nonetheless, the framework of a public good still provides an accurate picture or at least a sufficiently close approximation of the incentives people face in many situations. For stronger deviations, though, the framework of club goods and common-pool resources can be used.

Club or toll goods are non-subtractable but excludable.[47] A walled garden, golf club, and toll roads are examples of club goods. It is easy for the gardening- or golf club members or drivers who paid the toll charges to enjoy a garden, golf club or road without diminishing the quality and quantity of those goods available to other club members.[48] It is equally easy to prevent the non-members from enjoying the club good benefits. Due to the wide array of possible sizes of a club (from a few people to supra-national entities) some goods, such as public defence or clean air, may be considered public goods at some levels of aggregation but are better understood as club goods at other levels. Think for example about the benefits of a supra-national entity such as the European Union. From the perspective of an EU citizen, many of them may be considered a public good, as effectively no one within the "club" of EU citizens can be excluded. However, they are still excludable as EU citizenship is only given to the citizens of the EU member states.

---

[45] Ibid., 644 et seq.
[46] See also N. Purtova, 'Health Data for Common Good: Defining the Boundaries and Social Dilemmas of Data Commons' in Samantha Adams, Nadezhda Purtova and Ronald Leenes (eds), *Under Observation: The Interplay Between eHealth and Surveillance*, vol 35 (Springer International Publishing 2017) <http://link.springer.com/10.1007/978-3-319-48342-9_10> accessed 17 April 2021, 182 et seq.
[47] J. M. Buchanan, 'An economic theory of clubs' (1965) 32 Economica, 1.

|  |  | *Subtractability of use* | |
|---|---|---|---|
|  |  | *High* | *Low* |
| ***Difficulty of excluding beneficiaries*** | *High* | Common-pool resources | Public goods |
|  | *Low* | Private goods | Club (toll) goods |

Figure 1: Four types of goods[49]

Common-pool resources (CPRs for short, but also called "commons") are subtractable but not easily excludable. Natural resources such as forestry, water basins or fisheries are typically named as examples of CPRs. Hardin was among the first who on an example of grazing lands articulated how common use of shared resources moved by self-interest and lack of communication led to overgrazing and ultimately resource deterioration. This has become widely known as the "tragedy of the commons".[50] According to Hardin, the tragedy can be avoided either by establishing a market through the allocation of private property rights and allowing their efficient exchange or through centralised management of the resource by the government.[51] Further field studies of the common-pool resources by Elinor Ostrom and her followers within neo-institutional economics demonstrated, however, that the commons did not have to end in tragedy and other regulatory options beyond the traditional dichotomy 'market vs state regulation' were possible, notably, collective resource governance.[52]

---

[48] Although the increase in the number of club members beyond a certain point may lead to deterioration of enjoyment of a good by club members. Think of an overcrowded toll road, or a walled garden attended by too many club members at a time, resulting in more noise, litter, damage to grass and other plants and overall less enjoyable experience of the garden per club member.
[49] Elinor Ostrom, 'Beyond Markets and States: Polycentric Governance of Complex Economic Systems' (2010) 100 American Economic Review 645

In addition to their subtractability and low excludability, another characteristic of the CPRs is their complexity. CPRs are "system resources, meaning that they comprise entire 'resource ecosystems', a combination of interrelated and interdependent elements that together form a common-pool resource."[53] Natural CPRs like fisheries typically have a two-fold structure: stock (e.g. a fishing pond and the population of fish inhabiting the pond) and benefits, or 'flow units', produced by the stock (e.g. individual fish).[54] But other CPRs may have a more complex structure.[55]

When a CPR resource is present, we speak of a CPR situation. When the CPR resource is not used sustainably, we deal with a CPR problem.[56] To help analyse the CPR problems and inform their solutions, Gardner et al. group the commons problems into two types: the problems of provision and of appropriation:[57]

---

[50] Garrett Hardin, 'The Tragedy of the Commons' (1968) 162 Science 1243
[51] Ibid. 1244 et seq.
[52] Michael D. McGinnis and James M. Walker, 'Foundations of the Ostrom workshop: institutional analysis, polycentricity, and self-governance of the commons' (2010) 143 Public Choice 293, 296.
[53] Purtova (n46) 183
[54] Ostrom (n49).
[55] E.g. see discussion of Hess and Ostrom in 4.2(iii) on ideas, facilities and artefacts in relation to scientific knowledge.
[56] R. Gardner, E. Ostrom, and J. M. Walker. 'The Nature of Common-Pool Resource Problems' (1990) 2 Rationality and Society, 335–358; Ostrom (n49).

> *[T]he access to* or *allocation of the benefits of the resource is problematic in the appropriation problem, whereas the provision problem has the preservation, quality and sustainability of the resource stock at the heart of it.*[58]

While traditionally the CPR framework has been applied to natural resources, its relevance has expanded to other contexts such as wildlife and animal species[59] considered "global commons", cyberspace,[60] world oceans, atmosphere, Antarctic,[61] knowledge[62] and privacy.[63] One of the attractions of using the CPR framework across contexts is undoubtedly the Institutional Analysis and Development ('IAD') framework that Ostrom and her followers developed to diagnose and solve problems of collective action for sustainable governance of a shared resource.[64]

To conclude the account of the economic classification of goods, it is important to acknowledge the role of technology in the characterization of a good. Technology impacts the general availability of something as an economic good. For instance, the deep seas, the atmosphere, the electromagnetic spectrum, or space could only be seen as objects satisfying human needs when the technology to capture their benefits became available.[65] Technological developments can also cause a fundamental change in the nature of the resource, e.g. "with the resource being converted from a nonrivalrous, nonexclusionary public good into a common-pool resource that needs to be managed, monitored, and protected, to ensure sustainability and preservation,"[66] thus shifting the resource from one category to another.

---

[57] Gardner et al., 346.
[58] Purtova (n46) 188, citing E. Ostrom, R. Garder, and J. Walker, *Rules, Games, and Common-Pool Resources* (The University of Michigan Press, 1994), 9.
[59] Michael D. McGinnis and James M. Walker, 'Foundations of the Ostrom workshop: institutional analysis, polycentricity, and self-governance of the commons' (2010) 143 Public Choice 293.
[60] Priscilla Regan, 'Privacy as a common good in a digital world' (2002) 5 Information, Communication & Society, 382-405.
[61] http://www.unep.org/delc/GlobalCommons/tabid/54404/.
[62] C. Hess, E. Ostrom, 'Introduction: An Overview of the Knowledge Commons' in C. Hess & E. Ostrom (eds), *Understanding Knowledge as a Commons: From Theory to Practice* The MIT Press 1965); Katherine Strandburg, Michael J Madison and Brett M Frischmann (eds), *Governing the Knowledge Commons* (Oxford University Press 2014).
[63] Katherine Strandburg, Madelyn Rose Sanfilipp and Brett M Frischmann (eds), *Governing Privacy in Knowledge Commons* (Cambridge University Press, 2021); D. Bollier, 'The Growth of the Commons Paradigm' in Charlotte Hess and Elinor Ostrom (eds) *Understanding Knowledge as Commons Hess* (MIT Press, 2007), 31.
[64] Ostrom, Elinor, 'Background on the Institutional Analysis and Development Framework: Ostrom: Institutional Analysis and Development Framework' (2011) 39 Policy Studies Journal 7 <https://doi.org/10.1111/j.1541-0072.2010.00394.x>

In the subsequent sections, we critically review economic perspectives on data as an economic good and interrogate to what extent and for which purposes those perspectives are useful or unproductive to guide governance efforts in the digital society.[67]

## 4      Conservative view: non-rival information and data, data as a club good

Given the close relationship between information and data and that data is defined through information as its digital representation, we review the economic literature both on data *and* information. The conventional economics accounts of information and data overwhelmingly agree that information and data are non-rival.[68] While there is some disagreement as to the excludability of information, making it either a public or a club good, data is generally considered excludable, making it a club good.

Varian considers information goods in general as public goods because exclusion is too costly and the information is "inherently nonrival" due to low costs of reproduction.[69] Stiglitz agrees:

---

[65] Hess and Ostrom (n 62), 10.
[66] Ibid.
[67] The authors would like to acknowledge the research of Dr Sebastian Dengler that was informative to our understanding of the neo-classical economic classification of goods and information as a good. All mistakes are ours.
[68] Note that some information goods, such as consulting and education, are considered rival and excludable (e.g. Roxana Mihet and Thomas Philippon, 'The economics of Big Data and Artificial Intelligence' in J. J. Choi and B Ozkan (eds), *Disruptive Innovation in Business and Finance in the Digital World* (Emerald Publishing Limited 2019) 29 <https://doi.org/10.1108/S1569-376720190000020006> However, these can be seen as services for provision of information rather than purely information.
[69] Hall Varian, 'Markets for Information Goods', Daft paper October 16 1998 < https://citeseerx.ist.psu.edu/document?repid=rep1&type=pdf&doi=d0bc1ba2b540d3ab74c6fa69e2c75e1f7f3787be> last accessed April 8 2023.

> [I]nformation [is] fundamentally different from other 'commodities.' It possesses many of the properties of a public good – its consumption is nonrivalrous, and so, even if it is possible to exclude others from enjoying the benefits of some piece of knowledge, it is socially inefficient to do so; and it is often difficult to exclude individuals from enjoying the benefits.[70]

According to Moody and Walsh, information is infinitely shareable and not depletable.[71] Newman joins the characterizations of information as non-rival because the marginal cost of distribution and reproduction is zero or very low. However, he argues, the excludability of information is not a static characteristic and is "determined by the feedbacks of previous public policy decisions, which shape information asset characteristics,"[72] making information a club good when existing regulations enable excludability and a public good when they do not. Duch-Brown et al make a similar point that the excludability of information depends on technical and legal intervention.[73] Ciuriak notes that although knowledge is temporarily excludable by innovative firms e.g. by means of patenting, it eventually passes to the public domain when patents expire.[74]

More recent economics literature aiming to contribute to discourses on digital economy rather than information governance generally focuses on data. Since data is understood as digital representations of information, in these later accounts, it also inherits its non-rivalrous nature. But, unlike information, data is always excludable. There are no deviations in this assessment among the economists. To name a few examples, Koutroumpis et al write: "Ideas, patents, and data are non-rivalrous in use, in that a single idea or datum may be usable by many individuals and replicated at low marginal cost."[75]

---

[70] Joseph E. Stiglitz, 'The Contributions of the Economics of Information to Twentieth Century Economics' (2000) 115 Quarterly Journal of Economics 1441, 1448.

[71] Daniel Moody and Peter Walsh, 'Measuring the value of information: An asset valuation approach' (1999) Paper to be presented at the Seventh European Conference on Information Systems (ECIS'99), Copenhagen Business School, Frederiksberg, Denmark, 23-25 June, 1999.< https://www.researchgate.net/file.PostFileLoader.html?id=56b608e95dbbbd76628b4582&assetKey=AS%3A326144877449217%401454770408208> Last accessed April 8 2023.

[72] Amraham L. Newman, 'What You Want Depends on What You Know: Firm Preferences in an Information Age' (2010) 43 Comparative Political Studies 1286,1288

[73] Nestor Duch-Brown, Bertin Martens and Frank Mueller-Langer, 'The Economics of Ownership, Access and Trade in Digital Data' (17 February 2017) <https://papers.ssrn.com/abstract=2914144> accessed 8 April 2023.

[74] Dan Ciuriak, 'The Economics of Data: Implications for the Data-Driven Economy" in "Data Governance in the Digital Age' in Centre for International Governance Innovation, *Data Governance in the Digital Age* (Centre for International Governance Innovation 2018), 14 et seq. available online https://issuu.com/cigi/docs/data_series_special_report last accessed 8 December 2021

According to Jones and Tonetti, data is nonrival since it is infinitely usable, and excludable since access to data can be blocked by technical means, such as encryption,[76] next to legal measures to enable exclusion from access to digital data. Farboodi and Veldkamp,[77] Varian,[78] Dosis and Sand-Zantman,[79] Duch-Brown et al,[80] and many others make the same observations about data. In other words, the non-rivalrous nature and excludability of data are broadly accepted as well-established truths and not questioned among economists.

Classification of data as a club good accurately reflects the characteristics of data at present. Data is excludable due to its digital nature and hence strong ties to a physical carrier, such as a hard drive, a server or a data centre. Even when data is stored in a "cloud" which suggests something ethereal, there is always physical hardware involved, and one can be excluded from access to that hardware. The current realities of data access are full of examples of data pools managed as walled gardens, or clubs, where access is granted, e.g. to paying customers while everyone else is excluded from data benefits. For instance, data of Facebook users is generally only available via a specific API to those who pay for access, and Google controls who can place third-party cookies via its Chrome browser and access the data of the browser users.[81] There are cases of web scraping where data is accessible to all without barriers. However, such barriers *can* be erected in the form of security measures and legal prohibitions, which make data excludable. Data is also generally not subtractable as the quantity or quality of data does not generally diminish with use. It is suggested that the NFT technology– through the use of blockchain – can make a digital object unique by preventing it from being copied or transferred more than once.[82] This makes the replication of data impossible or very costly. Thus under some circumstances, data can become rival or subtractable and hence can also become a private good. However, making data not transferable is not a dominant tactic in the data economy at the moment, and in most cases, data remains a club good.

---

[75] Panetelis Koutroumpis, Aija Leipinen and Llewellyn D. W. Thomas, 'The (Unfulfilled) Potential of Data Marketplaces' (2017) ETLA Working Papers, No. 53, The Research Institute of the Finnish Economy (ETLA), Helsinki <https://www.etla.fi/en/publications/the-unfulfilled-potential-of-data-marketplaces/> last accessed April 8 2023.

[76] Charles I. Jones and Christopher Tonetti, 'Nonrivalry and Economics of Data' (2020) 110 American Economic Review 2819.

[77] Maryam Farboodi and Laura Veldkamp, 'A growth model of the data economy' (2021) Working Paper 28427 http://www.nber.org/papers/w28427> last accessed 4 December 2021, 2

[78] Hal Varian, 'Artificial Intelligence, Economics, and Industrial Organisation' in Ajay K. Agrawal, Joshua Gans, and Avi Goldfarb (eds), *The Economics of Artificial Intelligence: An Agenda* (University of Chicago Press 2018).

[79] Wilfred Sand-Zantman and Anastasios Dosis, 'The Ownership of Data' (2019) The University of Toulouse, unpublished manuscript <https://www.tse-fr.eu/publications/ownership-data> last accessed 6 December 2021.

[80] Duch-Brown, Martens and Frank Mueller-Langer (n73).

The chief criticism of the conservative view on data as a club good is that it considers data in isolation from its context: where the data comes from and what impact its extraction and further processing have on society. Data is a function of human activities, and their environments are increasingly mediated and captured through digital technologies. Once harvested, the data shape those activities and environments through nudging, algorithmic decision-making and other forms of algorithmic governance.[83] The conservative economic theory disregards this context. This narrow view is akin to considering fish sold on the market as a purely private good and in isolation from how it was caught, if it is an endangered species and the possible impacts of fishing on the biodiversity, preservation of the ocean ecosystem or local fishing communities. Such an account would not be inaccurate but also not complete.

---

[81] Although Google intends to exclude third-party cookies altogether in 2023 ("Google tracking cookies ban delayed until 2023" BBC News published on 25 June 2021 https://www.bbc.com/news/technology-57611701 last accessed April 8 2023.

[82] NFT (non-fungible token) technology makes it possible to sell digital objects as unique. For instance, the first Tweet of Jack Dorsey was sold as an NFT (Elizabeth Howcrift "Twitter boss Jack Dorsey's first tweet sold for $2.9 million as an NFT" Reuters, March 22, 2021 <https://www.reuters.com/article/us-twitter-dorsey-nft-idUSKBN2BE2KJ >, last accessed 20 December 2021.

[83] Shoshana, Zuboff, 'Big Other: Surveillance Capitalism and the Prospects of an Information Civilization' (2015) 30 Journal of Information Technology 75 <https://doi.org/10.1057/jit.2015.5> last accessed April 8 2023.

Some of the avant-garde conceptualizations of data as (a part of) the commons discussed below aim to address this shortcoming of the conventional economics classification.

Looking at data and governance of digital society through the prism of a club good has another significant limitation. The rationale of the economic classification of goods along the axes of subtractability and excludability is to recommend governance strategies that would ensure the provision of a sufficient quantity and quality of that good. For instance, market and private property have traditionally been seen as the appropriate governance mechanisms to create incentives for the production of private goods because the efforts of producing and maintaining private goods are rewarded when a private good is sold or otherwise exclusively enjoyed. To illustrate, the general consumer and other goods (private goods in economics terms) deficit in the USSR has been ascribed to the absence of a market economy and private property in the country.[84] Privatisation of state property and creation of market economy combined with political reforms in the late USSR and post-soviet Russia supposedly created incentives for the provision of the consumer goods and eventually solved this deficit. At the same time, public goods such as street lighting or a lighthouse signal are best provided by the state and would fall prey to free-riding or not be provided at all if left to the market forces, since lack of excludable benefits does not create incentives for the members of the public to contribute.

Following this logic, if the purpose of one's analysis is to find ways to ensure that sufficient quality and quantity of data is available, e.g. to enable innovation or generate wealth, one needs to focus on data and its characteristics as an economic resource to construct the governance scheme that would create incentives to create data production or availability. For instance, data possesses the characteristics of a club good, i.e. it is non-subtractable and excludable. The ability to exclude and exclusively benefit from data creates enough incentives to create the data at least for the actors of the so-called "data industry", e.g. by capturing consumers' behaviour to build an effective search engine or behavioural advertising ecosystem. Although data is a club good, it does not seem to be prone to the congestion problem as some other club goods, e.g. toll roads, are. So no regulatory intervention seems necessary to incentivize data production or prevent congestion. Some authors point out that the problem is not so much with data creation as with data sharing or pooling: while sharing data results in economies of scope and scale (aggregated data may be used for new purposes by new data holders and bigger data pools will supposedly render more value, e.g. for training AI), the costs of sharing may outweigh the benefits, especially for the data holders whose activities do not depend on the availability of data.[85] Think of hospitals. They routinely generate electronic health records, and do not immediately benefit from data analysis but will incur additional costs of infrastructure, legal compliance and labour to provide access to their data to someone else. For others whose business models are directly tied to data analysis, the disincentive might be that once the data is shared, even with a limited "club" of users, the benefits of data use are not excludable, i.e. all club members are able to extract the same insights from the same data pool, which might undermine one's competitive advantage. A good illustration is Google which is not sharing its search data with competing search engines voluntarily. Regulatory intervention is necessary to eliminate those costs and disincentives.[86]

---

[84] Marie Lavigne, *The Economics of Transition: From Socialist Economy to Market Economy* (2nd ed. Palgrave MacMillan 1999) 92.

At the same time, this "data as an economic good" frame of reference is not productive if one's objective is not to ensure the production or availability of *data*, but something else. Attempting to propose solutions to societal problems – even connected to data – by applying the economic goods analysis to data is fated to miss the target because this type of analysis is wired to produce solutions for the production or availability of *data* and nothing else. Therefore, if one wishes to use the economic goods frame of reference, e.g. to achieve the provision of privacy, democracy or other "goods" (since economics defines a "good" broadly), one needs to redefine what the good in question is.

---

[85] B. Carballa-Smichowski, N. Duch-Brown, and B. Martens, 'To pool or to pull back? An economic analysis of health data pooling' (2021) JRC Digital Economy Working Paper Seville: European Commission < https://joint-research-centre.ec.europa.eu/system/files/2021-12/jrc126961.pdf> last accessed April 8 2023.
[86] Ibid.

**5      Data as a common good: the landscape of the data commons**

In addition to the conventional economic classification of data as a club good, there is a more avant-garde and flourishing body of scholarship, which could loosely be labelled "data commons". While the literature in this strand is very diverse, all of it highlights some collective dimension of information or data and advocates for collective information or data management. In one way or another, the data commons literature rests on or responds to the commons analytical framework of Ostrom for sustainable management of the Common-Pool Resources. We roughly divide this literature into five strands. This classification might be imperfect, and some authors may not neatly fit under just one of the strands. Yet, this classification reflects the principal uses of the framework of the commons in relation to data and introduces some order to the perplexity of the data commons scholarship that has exploded in recent years.[87] Precisely because of the new-found popularity of the data commons, describing – even briefly - and reflecting on these literatures will take considerably more space in our overview compared to the neo-classical economic classifications. The five strands of the data commons scholarship are:

(1) Naturalist approaches to data commons,

(2) Information- or data commons for broader societal goals,

---

[87] The perplexity is reinforced by a linguistic overlap. Both philosophy and economics use the terms "public good" and "common good", but in very different meanings (Hussain Waheed, 'The Common Good' in Edward N. Zalta (ed) *The Stanford Encyclopedia of Philosophy* (2018) <https://plato.stanford.edu/archives/spr2018/entries/common-good/> last accessed 15 December 2022.

(3) Governing the knowledge commons,

(4) Commons-based peer production, and

(5) Relational commons.

The limited scope of this paper does not allow for a detailed review of each of these strands. Still, we will briefly introduce them here with three questions in mind: 1) what resource do these authors see as a CPR, 2) what subtractability and sustainable resource use mean for these authors, and 3) what governance strategies they suggest. We will follow with reflections on their value and shortcomings.

The commons-based peer production and relational commons strands do not strictly meet the selection criteria as they do not focus on the excludability and subtractability of data or other resources (and hence move outside of the matrix as presented in Figure 2). Yet, without them, the account of the commons scholarship would be incomplete and we include them in the selection.

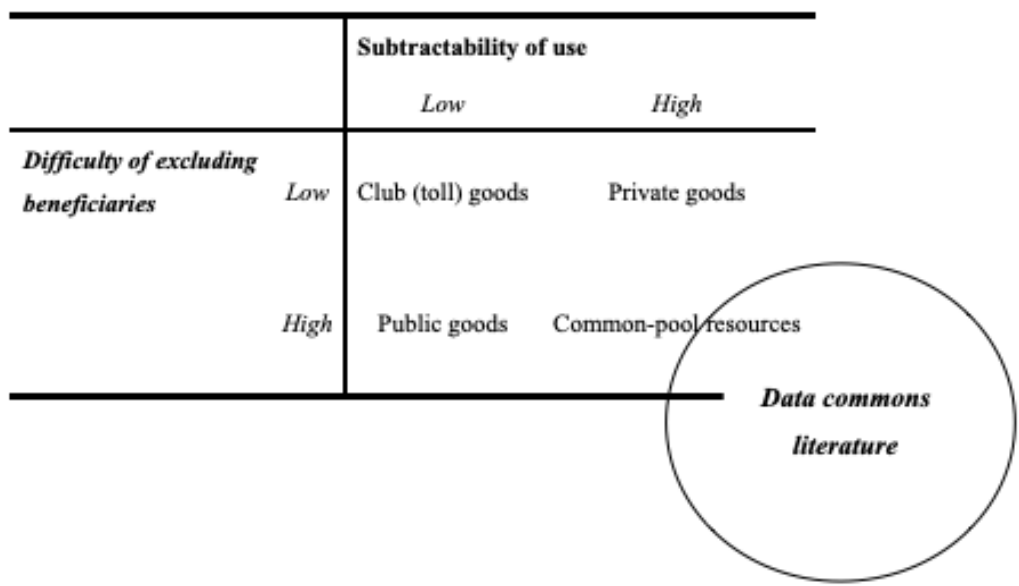

Figure 2: Relationship between the neo-classical economic classification of goods and the commons literature

## 5.1 Naturalist approaches: subtractable data as a common-pool resource

The naturalist strand of the data commons scholarship is a literal application of Ostrom's CPR framework to data based on the idea that data by its nature is subtractable and difficult to exclude from and hence a common-pool resource.

Regan argued that personal information (the US functional equivalent of "personal data" in the EU) is a common-cool resource because it possesses three features: 1) it is available to more than one person; 2) it is difficult to exclude users from it, and 3) it is subject to "degradation as a result of overuse".[88] According to Regan, the quality of the personal data flow is of value to many but will degrade with use. Personal information will become "[i]ncomplete, inaccurate and/or outdated" as data subjects will "resent or distrust this market in secondary uses of personal information" and withdraw from it.[89] Information may lose value to one information appropriator by its use by another appropriator.[90]

Similarly, Zygmuntowski, Zoboli and Nemitz[91] argue that "[i]t is not collective management or sharing … that constitutes data as [a] CPR, but the nature of data."[92] Technological developments, in particular expansion of surveillance apparatuses, have changed data which before was replicable and non-rival, into a subtractable good, where subtractability refers both to the negative consequences of data use for individuals and society, but also to undersupply, pollution, lack of quality and findability of data itself.[93]

The natural consequence of qualifying data as a CPR both for Regan and Zygmuntowski et al is that data should be governed according to Ostrom's design principles. Regan argues for the collective governance of personal information that is 1) "support[ed] by higher authorities in applying sanctions"; 2) access to the resource system is clearly defined; 3) clear resource boundaries are established; 4) resource users participate in devising rules; 5) "graduated sanctions [are created] for offenders"; and 6) low-cost conflict resolution mechanisms exist.[94] Zygmuntowski et al advocate for "public data commons, defined as a trusted data sharing space established in the public interest" aimed at "safeguarding … European values and rights by active participation of public actors in stewarding data".[95]

---

[88] Regan (n60), 392 et seq.
[89] Ibid. 393.
[90] Ibid.
[91] Zygmuntowski, Zoboli and Nemitz (n13).
[92] Ibid. 16.
[93] Ibid. 15-16

---

[94] Regan (n60) 401.
[95] Zygmuntowski, Zoboli and Nemitz (n13) 18.

## 5.2 *Information- or data commons for broader societal goals*

This strand of the scholarship applies Ostrom's CPR framework to information and data in a less literal sense. They do not claim that data or information itself are a CPR and that they deteriorate with use. In fact, something else (scientific knowledge, data ecosystems) is a common-pool resource and this is the sustainability of that something else that they are concerned about. Hess and Ostrom herself, Purtova and Wong et al are all illustrative of this interpretation of the commons in the information- and data context.

Hess and Ostrom were the first to apply Ostrom's CPR analytical framework to information, knowledge and data.[96] They identify scholarly information and knowledge as a common-pool resource. Similar to the natural CPRs, scholarly information is a complex resource. Yet, its "anatomy" is different from the "stock and flow units" structure of the natural CPRs. Hess and Ostrom propose a three-way distinction between artefacts, facilities, and ideas, or what they call "the ecological makeup of scholarly information".[97]

An *artefact* is an observable and nameable, physical or non-physical representation of an idea[98] such as digital or printed books and articles. Physical artefacts are excludable, but replenishable. A *facility* is where the artefacts are stored and made available. Think of a library or an archive, but also the Internet. They are excludable, and many facilities such as libraries and archives have well-established policies of access. They are subtractable as they are subject to deterioration if not maintained.[99] The *ideas* are the "intangible content" of the artefacts, including data, "innovative information and knowledge". They are "non-physical flow units".

---

[96] Charlotte Hess and Elinor Ostrom, 'Ideas, Artifacts, and Facilities: Information as a Common-Pool Resource' (2003) 66 Law and Contemporary Problems 111.
[97] Ibid. 129
[98] Ibid.
[99] Ibid. 129-130.

While the use of an idea by one person does not subtract from it, it is possible to exclude from an idea, e.g. by keeping it secret.[100]

For Hess and Ostrom, scientific knowledge is subtractable as it deteriorates as a result of unsustainable management. Digitization caused a substantial change in the ecology of scholarly information. While in the pre-digital age libraries played a key role in scholarly communication and provided easy access to physical books and journals, digital publication gave publishers, a.o. through intellectual property rights, control over the availability of artefacts to the libraries and scholars. Particularly smaller libraries and their users have been restricted from access to scientific knowledge as a result.[101] Hess and Ostrom suggest that the shift to the digital forms of scientific communication with the big publishers controlling access to ideas made the scientific knowledge subtractable, i.e. subject to deterioration. The publisher-controlled mode of scientific communication restricts smaller libraries and their users, often from poor communities, from access to scientific knowledge (appropriation). At the same time, restricted access to the existing scientific knowledge logically inhibit the production of new ideas to the impoverishment of the knowledge commons (provision).

In line with the IAD framework for CPRs, Hess and Ostrom suggest that collective governance is a potentially successful strategy to ensure sustainability of scientific knowledge. They observe that successful collective scientific information management initiatives have emerged to counter the exclusion, leading to the rise of new institutions enabling scientific communication, such as self-archiving and open access movement.[102]

---

[100] Ibid. 130.
[101] Ibid. 134 et seq.
[102] Ibid. 143.

Purtova's commons analysis of governing digital society and (personal) data closely follows Ostrom's work.[103] Two points in Purtova's analysis stand out the most: she identified the "data ecosystem" as a CPR and further develops Ostrom's ecological approach to the commons governance in the digital context and offers a broad interpretation of subtractability beyond deterioration of data which makes the CPR framework applicable to a wider range of situations and problems .

The starting point of Purtova's analysis is identifying what a CPR at hand is. It cannot be health- or personal data, as both these categories are too dynamic (any personal data can become health data and any data can become personal) while successful CPR governance requires stable resource boundaries.[104] Purtova proposes to consider 'data ecosystems' as a CPR instead.[105] This is because "personal data does not exist in isolation and presents itself as, or forms a part of, a system resource" composed of three elements: people, platforms and data.[106] Within the 'data ecosystem', "people and data inherent in the very fact of [their existence] are the core resource, and the data collected about or in relation to them is simply a benefit [flow unit] generated by the core resource."[107] This ecosystem approach "allows for simultaneous existence and interaction of multiple [data] ecosystems of various sizes and levels, that do or do not overlap, consume smaller ecosystems and are consumed by larger ecosystems."[108]

Data commons construed in such a way is subtractable in a sense different from the physical exhaustion or corruption of data:

---

[103] Purtova (n46).
[104] Ibid. 189 et seq.
[105] Ibid. 195.
[106] Ibid. 195.
[107] Ibid. 196.
[108] Ibid. 197.

> [T]he data commons sustainability dilemma … instead should be understood in terms of the long-term effects of commoditization of personal data and modern data processing practices, compromising [the] survival of certain social values and hence leading to 'extinction of society' as we would like it to be, taking the shape of 'data poaching'.[109]

In other words, the CPR problems of the data commons are altering the fabric of society (the provision problem) and data enclosure by big tech (the appropriation problem).[110]

This particular paper only "aims to redirect the focus of the … data commons debate [from data sharing] to the social dilemmas and to the politics of data sharing"[111]. In subsequent work, Taylor and Purtova argue that Ostrom's IAD principles can be used as a roadmap to sustainable data commons governance.[112]

Wong et al.[113] is another recent attempt to apply Ostrom's CPR framework to further societal goals beyond prevention of deterioration of data. It is clearly inspired by Ostrom, but uses the CPR framework selectively. Notably, it does not clearly identify a resource that is a CPR or explain what makes it subtractable, but does suggest CPR-inspired governance strategies.

Wong et al. version of data commons is aimed at furthering data protection in contrast to the data commons practices "prioritiz[ing] data sharing, data curation, and reuse".[114] Wong et al advocate for commons as "consensus conference[s]" that "encourage dialogue among data subjects, experts, and policy-makers and ordinary citizens, creating new knowledge together for the common good".[115] The commons as a data governance method is combined with ideas about the value of collective discussions, knowledge-sharing, and decision-making about data, which will help individual members of the commons to make better decisions about their 'data protection preferences'.[116]

---

[109] Ibid. 200.
[110] Ibid. 200, 203.
[111] Ibid. 180.
[112] L. Taylor and N. Purtova, 'What is responsible and sustainable data science?' (2019) 6 Big Data & Society <https://doi.org/10.1177/2053951719858114>.
[113] Janis Wong, Tristan Henderson, and Kirstie Ball, 'Data Protection for the Common Good: Developing a Framework for a Data Protection-Focused Data Commons' (2022) 4 Data & Policy <https://doi.org/10.1017/dap.2021.40>.
[114] Ibid. 11 (e.g. medical data pooled for medical research).

## *5.3  Governing the knowledge commons*

Another notable approach towards the governance of information goods including data is the "governing the knowledge commons" framework ('GKC'). It has been authored by Frischmann, Madison, and Strandburg and applied by numerous other scholars in contributions to a series of books edited by the GKC intellectual founders.[117]

While having its intellectual roots in Ostrom's work, the GKC authors present the GKC as an updated and improved version of Ostrom's IAD framework.[118] They argue that the original IAD framework is better suited for the governance of the natural CPRs but not for the intangible resources such as knowledge, information, and data,[119] as these are not excludable or depletable, and hence are not CPR's.[120]

The GKC turns around the relationship between resource and its governance: while for Ostrom the characteristics of a common-pool resource (difficulty to exclude and subtractability) dictate the best governance models (collective institutions), for the proponents of the GKC framework the collective governance makes a resource a commons and the institutions determine the characteristics of the governed resource. Put differently, what makes a resource, including data, [121] a commons is the fact that it is governed by a collective.[122] This makes the GKC framework suitable for a broader range of goods and governance contexts, including privacy[123].

---

[115] Ibid. 8.
[116] Ibid. 12-13.
[117] Frischmann, Madison and Strandburg (n62); Katherine J. Strandburg, Brett M Frischmann, and Michael J. Madison (eds), *Governing Medical Knowledge Commons* (Cambridge University Press 2017); Sanfilippo, Rose and Frischmann (n62).
[118] While the GKC authors explicitly acknowledge their departure from the IAD principles, they in fact also depart from the entire CPR framework which gave rise to those principles. This is manifest in their approach to qualify something as a commons based on its collective governance and not its natural characteristics.
[119] Madison, Michael, 'Tools for Data Governance' (2020) Technology and Regulation 29 <https://doi.org/10.26116/techreg.2020.004> 35. Last accessed April 8 2023.
[120] Ibid. 35; See for a more extended argument on the limitations of applying IAD to knowledge, Brett M Frischmann, Michael J Madison and Katherine J. Strandburg, 'Introduction' in Brett M. Frischmann, Michael J. Madison and Katherine J. Strandburg (eds) *Governing the Knowledge Commons* (Oxford University Press 2014), 17.

The subtractability of any resource managed in common also plays no role within the GKC frame of thinking. Instead, the collectives are "to define [their] own governance system relative to dilemmas associated with specified resources, producing a form of institutional governance in context."[124]

While taking a collective's institutional (self)governance as a starting point, the GKC authors do not put forward prescriptive guidance on *how* resources should best be governed. The GKC framework merely "supplies a means of describing the breadth of the field in a systematic way."[125] The study of the commons through the GKC is "embodying a set of strategies that solve coordination problems [of collective action]".[126] This is done with the help of a set of questions to be asked about the studied commons that accompany the framework.[127] This makes the GKC framework primarily a descriptive tool to be used for researchers studying (data) commons empirically, rather than a theory of how best to engage in the governance of resources in common.

---

[121] Madison (n119).
[122] Ibid. 31; Frischmann, Madison and Strandburg (n120) 17.
[123] Sanfilippo, Frischmann, and Strandburg (n62).
[124] Madison (n119) 36.
[125] Ibid. 36.
[126] Ibid. 35.
[127] Frischmann, Madison and Strandburg (n120); Brett M. Frischmann, Michael J. Madison, Katherine J. Strandburg, 'Conclusion' in Brett M. Frischmann, Michael J. Madison, and Katherine J. Strandburg (eds) *Governing Knowledge Commons* (Oxford: Oxford University Press 2014).

## *5.4 Commons-based peer production*

Recent data commons literature often draws on the information commons (in the sense of open culture, information and knowledge) and peer production discourse of the late 90s-early 00s which is well illustrated by the free and open software movement. The work of Benkler advocating for the 'commons-based peer production' (CBPP) of information[128] has been especially influential in the data commons literature.[129] We have not come across a version of the data commons argument based primarily on Benkler's analysis. Literature from all four remaining strands is thickly sprinkled with Benkler references, but only a selection of his arguments is often used.

This vision of the commons does not strictly fit the selection criteria as it does not focus on data or any other common-pool resource and does not explore their excludability and subtractability. Contrary to the Ostrom-influenced accounts of the data commons considered earlier, Benkler understands information (and information goods) as a public good which is nonrival and has zero (re)productive costs.[130] The commons enter the picture here to reflect and make a normative argument in favour of the collective access to and peer production of those goods.

According to Benkler, in the digital economy the production of information goods such as software, books, ideas, etc. through CBPP is more efficient, and hence desirable, in comparison with market or hierarchy-premised types of production.[131] Precisely because those goods are nonrival, they can be shared with zero reproduction costs[132] and the availability of information stimulates its reuse for the production of new information.[133] Therefore the free and open licensing formats of information should be stimulated.

---

[128] Yochai Benkler, 'Freedom in the Commons: Toward a Political Economy of Information' (2003) 52 Duke Law 1254.
[129] E.g. J. Birkinbane, *Incorporating the Digital Commons: Corporate Involvement in Free and Open Software* (University of Westminster Press 2020); Michael Bauwens, Vasilis Kostakis, and Alex Pazaitis, *Peer to Peer: The Commons Manifesto* (University of Westminster Press 2019); M Dulong de Rosnay and F Stalder, 'Digital Commons' (2020) 9 Internet Policy Review <https://policyreview.info/concepts/digital-commons> accessed 24 November 2021; Zygmuntowski, Zoboli, and Nemitz (n13); Tommaso Fia, 'An Alternative to Data Ownership: Managing Access to Non-Personal Data through the Commons (2020) 21 *Global Jurist* 181–210; and many others.
[130] Benkler (n128) 1252.

Further, CBPP of information goods is conducive to a variety of social-democratic values, notably, democracy, individual autonomy, and social justice. Democracy, for instance, is promoted through the increase in the number of individuals capable of participating in the collective action. In addition, grass-roots peer production of newsworthy information decentralises and diversifies knowledge production which does not have to rely solely on a handful of information intermediaries as a result.[134]

An important part of Benkler's analysis is the emphasis on the need to think beyond the resources that are produced as a result of the CBPP. Doing commons-based peer production necessitates the core common infrastructures – sets "of resources necessary to the production and exchange of information, which will be available as commons"[135]. These infrastructures might be created by, for instance, removing cumbersome market and hierarchy-oriented legislation that prevent the free distribution of information.

---

[131] Note here the fact that he thinks that market, state, and commons are comparable in that way (Antonios Broumas, *Intellectual Commons and the Law: A Normative Theory for Commons-Based Peer Production* (University of Westminster Press 2020). This is a particular ('functionalist') way of comparing the value of the commons vis-a-vis the state and market. As we shall explain below, the commons are sometimes understood to be more than a mere management or governance solution and hence not to be evaluated on the basis of their capacity to solve a governance problem.
[132] Benkler (n128) 1252.
[133] Ibid. 1253.
[134] Ibid. 1262-63.
[135] Ibid. 1273.

## 5.5   *The relational commons*

This strand of data commons literature is characterized by considering data and its governance in terms of their relationship to and impact on the communities.[136] For conceptual clarity, it is useful to disentangle their analysis of the commons problem and the solutions they propose.

Dulong de Rosnay and Stalder understand the digital commons as "a subset of the commons, where the resources are data, information, culture and knowledge which are created and/or maintained online."[137] They understand the commons themselves as "a plurality of people (community) sharing resources and governing them in their own relations and (re)reproduction processes through horizontal doing in common, commoning".[138] Instead of seeing the commons as the management or the governance of a resource or a resource system, the *relationships* and the *dependencies* are defining.[139] The commonsnare better described as a 'life-form' than the management of goods or resources. It matters more what kind of community is related to a certain resource, and in what way, than the specific governance model or set of rules to be applied to that resource.[140] Goods like data are important for the long-term viability of the community, and not for the price one would pay for them on the market (or their 'exchange-value').[141]

---

[136] Parminder Jeet Singh, 'Data and Digital Intelligence Commons (Making a Case for Their Community Ownership)' (2019) SSRN Scholarly Paper 3873169. Rochester, NY: Social Science Research Network. <https://doi.org/10.2139/ssrn.3873169>.

[137] Dulong de Rosnay and Stalder (n129) 2.

[138] Ibid. 2. They quote here Massimo De Angelis, *Omnia Sunt Communia: On the Commons and the Transformation to Postcapitalism* (Zed Books 2019) 10.

[139] David Bollier, *Thinking like a Commoner* (New Society Publishers 2014) 15: "They come to depend on each other and love this forest or that lake or that patch of farmland. The relationship between people and their resources matter."

[140] Ibid. 15: "… commons are not just things or resources. … but they are more accurately defined as paradigms that combine a distinct community with a set of social practices, values and norms that are used to manage a resource. Put another way, a commons is a resource + a community + a set of social protocols. These three are an integrated, interdependent whole."

[141] Ibid. 108.

Similar to the indigenous data sovereignty,[142] community data rights[143] and other literatures, this version of data commons is critical of data commodification. Relational commons reject the market and capital as dominant mode of valuing goods like data, because those markets tend to transform the human relationships into goods and commodities.[144] The 'use-value' that the communities attach to the data is obscured and degraded by the 'exchange-value' attached to the data in market exchange.[145] Treating data as a good or a resource is frowned upon in this strand of literature.[146] Therefore, like the CBPP, the relational data commons fall outside of the economics goods matrix as represented in Figure 2.

Lijster is one of the few authors who proposes the governance strategies based on the idea of relational commons specifically for data. He argues for doing (data) commons as a political project.[147] This means that the commons efforts should be guided by a specific normative principle that explains and justifies the political direction (digital/data) commons head towards.[148] In particular, commons must try to limit and mitigate the implications of 'data capitalism'.[149] This can be done by communities taking back control of the resource and infrastructure from the big tech that endanger the commons as ideal and practice.[150] Data and platforms are at least part of why we have become dependent on big tech and should be controlled and owned by the commons. These should include everyone who is being exploited or extracted in contemporary digital capitalism.[151] Precisely because "the 99% of the population" have become the input or resources to be used in the capitalist production processes, the commons as a political project cannot be limited to small communities.[152] The commons should aim at the control of the 'big' globally active platforms.

---

[142] Maggie Walter, Tahu Kukutai, C Russo Carroll, and Desi Rodriguez-Lonebear (eds) *Indigenous Data Sovereignty and Policy* (Routledge 2021).
[143] Singh (n136).
[144] D Bollier and S Helfrich, *Free, Fair and Alive: The Insurgent Power of the Commons* (New Society Publishers 2019) 54-55; Bollier (n139) chapter 10;
Bollier and Helfrich 2019, 54-54; Bollier (n139) chapter 10; T Lijster, *Wat We Gemeen Hebben: Een Filosofie van de Meenten* (De Bezige Bij 2022) 108.
[145] The problem thus is that data is being used differently, when seen as removed from a real-life context.
[146] Bollier (n139) chapter 10; see also Bollier and Helfrich (n144) 54-54 on the concept of resource and 165-166 for that of the commodity.
[147] Lijster (n144).
[148] Ibid. 84-85.
[149] Ibid. 92-94; on data capitalism, see SM West, 'Data Capitalism: Redefining the Logics of Surveillance and Privacy' (2019) 58 Business & Society 20.
[150] "Precisely due to the growing dependence [on big tech], we should understand the data and platforms they use and manage, as commons, and for this reason find ways to reclaim control over them" (Lijster (n144), 133. Translation is ours).

## 6  How can data commons literature be productively used?

What do we make of this rich body of the data commons literature and how can it be productively used? The analysis below focuses on three points: when it makes sense to speak of the data commons, the strengths of the data commons analysis, and its limitations.

### *6.1  When the data commons analyses can be used.*

Our review of the data commons literature revealed that the commons analysis has been applied to data based on four kinds of claims: 1) an empirical claim that data possess characteristics of a common-pool resource and hence lends itself particularly well for the collective governance;[153] 2) an empirical claim that not data itself but another data-related good is a common-pool resource and should be governed as such to ensure its sustainability,[154] 3) an empirical claim that when data is governed in common it becomes the commons,[155] and 4) a political-moral claim that data ought to be commons and governed in common.[156] The validity of the empirical claims and compatibility of the moral-political claims with the moral-political context of the governance situation will largely determine whether or not a particular strand of the data commons literature can be productively used to inform data-related governance strategies.

---

[151] Ibid 192.
[152] Ibid 192.
[153] Literature considered under "Naturalist approaches".
[154] Literature considered under "Information- or data commons for broader societal goals".
[155] Literature from the GKC strand.
[156] Literature from the CBPP and relational commons strands.

*Data is not a common-pool resource and hence the naturalist version of the data commons cannot be used productively*

We submit that the first empirical claim that data itself is a common-pool resource in the sense of neo-classical economics is not valid and hence the naturalist data commons analyses should be rejected as unproductive. There is nothing in the nature of data as an economic good that calls for its sharing or collective governance. The naturalist data commons analyses do not convincingly dispute the classification of data as a club good, i.e. as a non-rival and excludable resource.[157] While the presence of negative effects of data practices on society and individuals is undeniable, contrary to what Zygmuntowski, Zoboli and Nemitz claim[158], they do not support a conclusion that the data itself is subtractable, i.e. diminishes in quality or quantity as a result of those practices.[159] Regan's prediction made in 2002 that predatory privacy-unfriendly data practices will result in people withdrawing from cyberspace and in diminishing quality and quantity of data[160] received no confirmation in practice. If anything, the quantities of data generated have only increased,[161] and withdrawing from cyberspace when essential elements of modern life such as social security, public transport, and healthcare are increasingly mediated by digital technologies is not a realistic option for a majority of people. Zygmuntowski et al who argue that the current data practices lead to poor data quality and limited data findability do not demonstrate convincingly the causal link between data practices and poor data quality. The fact that less data is available (or "findable") because of the capture by Big Tech does not mean that less data is being or can be generated. It merely shows that a few powerful actors have been effective in excluding others from the data, which supports a conclusion that data is a club good (easily excludable) rather than that it is by its nature subtractable and hence a common-pool resource akin to e.g. fisheries. In addition, similarly to the limitations of the neo-classical analyses discussed in 4.1, the data-centric commons analysis (i.e. "data as a common-pool resource or common good") is not productive if one's objective is not to ensure the production or availability of *data*, but other societal goals, like protection of privacy that both versions of the naturalist data commons that we considered advance. Therefore, we reject the naturalist version of the data commons as internally inconsistent and hence unproductive.

---

[157] See 4.1.
[158] Zygmuntowski, Zoboli, and Nemitz (n13).
[159] Ibid. acknowledge this as well at 15-16.
[160] Regan (n60) 393.
[161] E.g. <https://www.businesswire.com/news/home/20200508005025/en/IDCs-Global-DataSphere-Forecast-Shows-Continued-Steady-Growth-in-the-Creation-and-Consumption-of-Data>

*CPR analysis can be applied to data as a part of a socially constructed commons*

While there is nothing in the nature of data as a resource that demands its collective governance or data pooling, data can be seen as a part of another common-pool resource which is subtractable and difficult to exclude from. The CPR framework can be applied to data in this context in order to ensure the sustainability of that other resource, as Hess and Ostrom, Purtova and Wong et al have done. The literature reviewed construed scientific knowledge, data protection, or the "society as we want it to be" as such CPRs, but one can think of other examples, such as the digital economy. In particular, this would justify applying the IAD principles and the evidence from Ostrom-inspired empirical studies of commons to construct the governance of that CPR.

      The range of situations where the Ostrom-inspired governance recipes can be used is quite broad because – while the classification of a resource as a CPR is based on empirical insights about that resource – what amounts to its subtractability and hence what a CPR is is socially constructed. For instance, Purtova defines subtractability "in terms of the long-term effects of commoditization of personal data and modern data processing practices [on] social values and […] 'extinction of society' as we would like it to be".[162] The physical deterioration

---

[162] Purtova (n64) 200.

of a (part of) a resource (e.g. extinction of society) is not required before a situation can lend itself to the CPR analysis and CPR-inspired solutions.

*Data commons analysis based on political-moral claims*

Finally, the CBPP- and relational data commons analyses based respectively on the political-moral claims that data ought to be pooled or governed in common will be productive for those who share these political-moral views, with some limitations.

The relational commons literature has not yet developed sufficiently to propose specific governance strategies for data, and where those governance strategies are proposed,[163] reconciling them with the current realities might prove very difficult. The idealistic view on the democratic and non-commercial governance of data (data should not be subject to market exchange, communities should take back control over their data from big tech) are hard to mix with the current reality of data practices and infrastructures that are capitalistic and have so far developed along the market-oriented path. Reshaping the data markets and infrastructures according to the "relational commons" mould is likely to be extremely challenging if at all possible. Even if this enterprise succeeds, one wonders if applying the relational commons language and toolbox to deeply problematic practices, like data extraction, commodification and surveillance, to name a few could in fact legitimise and "commons-wash" those.[164] In other words, simply moving data-related problems into the domain of collective governance will only achieve collective governance and not necessarily provide solutions to the problems.

As for Benkler's commons-based peer production, we have significant reservations about how useful it is in the data-related contexts. We see at least three difficulties with using

---

[163] E.g. Lijster (n144)
[164] M. Dulong de Rosnay, 'Commonswashing – A Political Communication Struggle' (2020) 2 *Global Cooperation Research - A Quarterly Magazine* 11.

Benkler's work to substantiate a moral-political claim that data should to be the commons, in the sense that it must be open and produced in common.

First, Bankler's ideas have emerged primarily in relation to 'culture', 'information', and 'knowledge'. It is therefore important not to presume but to question to what extent knowledge and data are to be treated similarly, and to what extent the moral-political values attributed to the opening of culture, information, and knowledge, apply equally to data[165] which is a related but still a different thing.[166] CBPP was premised on specific (and optimistic) ideas about how the market, the public sphere, and public institutions function, and that the sharing of certain informational goods could foster specific social-democratic values. These are empirical presuppositions that cannot be relied on in other contexts, such as data sharing, without their validity in those contexts being verified first.

Second, even if we accept that Benkler's thinking about culture, information and knowledge can be applied to data, there is a risk that it will result in a data-centric regime wired for data production, without the added benefits of creating the open culture that is often associated with the sharing and open access of information and knowledge.

Finally, peer production is a mode of production where specific participants ('peers'), with specific skills and interests, work voluntarily on the production and dissemination of specific goods. Even *if* it is assumed that the market and public sphere function like Benkler assumes, and even *if* CBPP produces the goods and values it promises to deliver, the special character of CBPP cannot be generalised and applied to all forms of data governance. Not everyone is a peer and has resources and skill sufficient to work with open data.

---

[165] Compare, Arwid Lund and Mariano Zukerfeld *Corporate Capitalism's Use of Openness: Profit For Free?* (Palgrave Macmillan 2020); N Tkacz, 'From Open Source to Open Government : A Critique of Open Politics' (2012) 12 *Ephemera: Theory and Politics in Organization* 386.
[166] See part 2.

*The GKC-commons analysis can be applied to all instances of common governance of data.*

The GKC-type data commons analysis can be applied to any instance of collective governance of information resources, including collective governance of data, since its applicability is not conditional on the empirical characteristics of a resource and does not rely on any political-moral claims. However, while Ostrom-inspired CPR analyses suggest the governance strategies to ensure the sustainability of a data-related resource, and the relational and CBPP-based data commons prescribe that data ought to be governed and produced collectively, the GKC version of the data commons does not prescribe when data should be governed in common or to what purpose. While for Ostrom the end-game of the collective governance of the CPR is the sustainability of that resource, the GKC framework is morally and politically 'neutral' and does not offer any value to replace sustainability and to guide the collective governance towards a specific normative outcome - apart from the internal legitimacy or independence of the collective governance itself. The question GKC prompts is if one can be neutral with respect to (data) governance. Obviously one *can*, but it is to be doubted whether one *should* when many have shown the potential damage done by problematic forms of data gathering, processing, or governance.[167] Process-oriented values present in the GKC - independence, legitimacy - might on their own be insufficient to help differentiate between good and bad (data) governance forms and substance. Also in the specific case of *data commons*, Madison does not offer strong normative recommendations about how, where, and by whom data should be governed in common.[168]

When the GKC-version of the data commons is so broadly applicable and attracts such a mass of studies on collective governance in various contexts, as the GKC did, it is bound to produce useful insights on how to structure collective action. The reverse side of such inclusiveness of the GKC is that it might diffuse the political power present in the commons claims over data based on alternative, more politically charged visions, like the commons-based peer production and relational commons.

---

[167] See for references 4.vi.
[168] Madison (n119) 42.

## 6.2 Strength of the data commons analyses: problem analysis and ecological thinking about data-related problems

Some accounts of the data commons are presented as 'the way out of surveillance capitalism which increases welfare without killing innovation,'[169] allowing you to eat your proverbial cake and keep it too: use the data while preserving privacy, public interest, or another interest that is usually at risk as a result of data use. This is likely a legacy of the origins of the commons analysis in the governance of natural resources where the CPR framework and IAD principles were proposed as the way to appropriate a resource at a sustainable rate without destroying it. While attractive, this framing is oversimplifying the value of the commons way of thinking about data. The core strength of the commons literature in our view lies in its problem analysis. What distinguishes the commons from other classifications of data as an economic good is the ecological thinking that acknowledges the complexity of the data-related problems and draws attention to the broader societal, technical and economic context of production and use of data in connection to broader societal goals. Data commons push us to think about data-related problems and solutions in terms that are beyond data. This feature is observable to some extent in all versions of the data commons we reviewed but is especially apparent in Ostrom-inspired analyses reviewed under "Information- or data commons for broader societal goals" which employ ecological thinking about resources to be governed and problems to be solved.

---

[169] Jan J. Zygmuntowski [@ZygmuntowskiJ], 'Is There a Way out of Surveillance Capitalism Which Increases Welfare without Killing Innovation? I Can Tell You: There Is. And It Will Change How You Think of Your #data. If You Want to See More Democracy in Modern Information Society, Our #DataCommons Primer Is for You 👇 Https://T.Co/V8bFvDSXvs' <https://twitter.com/ZygmuntowskiJ/status/1580458137718775809> accessed 12 April 2023.

Unlike the conventional economic accounts, Hess and Ostrom recognize the complexity of the information goods and that the sustainability of an information-related good (in this case, scientific knowledge) does not depend solely on the provision of data but is in fact a function of the interaction of multiple parts forming an ecosystem. Before considering the governance strategies for an information-related good, one needs first to make a normative judgement on what good is desirable and needs to be sustainably managed (what is a CPR?) and then map the "ecological makeup" of that good: what elements and actors are essential for its sustainability. While the ecological makeup of the scientific knowledge commons (ideas, facilities and artefacts) is unique to the scientific knowledge commons, the same ecological thinking can be applied to other data-related goods. For instance, one can be inspired by Hess and Ostrom to construe a model where digital economy, privacy or other desired "good" is a CPR, move to map its ecological makeup where data may or may not play a key role, and think of the governance strategies based on that makeup.

Purtova seems to do exactly this. She argues that data itself is not the CPR but a part of a larger, more complex resource ecosystem. She further develops Hess and Ostrom's idea of the "ecological makeup" and proposes the notion of the data ecosystem which highlights "the interconnectedness of and the relationship between the elements of the system resource."[170] She argues that "personal data does not exist in isolation and presents itself as, or forms a part of, a system resource" composed of three elements: people, platforms and data.[171]

The data commons literatures that do not follow Ostrom emphasise the complexity of data-related problems and the connection of data to society as well, e.g. through the impact of the data use, production or governance on the communities or society. Madison, a key GKC author, advocates for thinking about data in terms of complex systems or ecologies.[172] The relational commons literature sees the impact (positive and negative) of data on communities as the defining element of the data commons. If Benkler's version of the CBPP is transplanted from the context of open culture and information to data,[173] the collective production and management of data will have impact on the state of democracy, if facilitated through creation of the public infrastructures, etc.

---

[170] Purtova (n64) 197.
[171] Ibid. 195.
[172] Madison (n119) 42.

In other words, the major takeaway of the data commons literature is that many problems that might be commonly associated with data are complex and a function of many variables. They are not necessarily exclusively data problems and will not be resolved by focusing the governance efforts of data alone. The first analytical step that needs to be made before designing a governance strategy is to define the problem and the objective which is rarely "how to govern the data" but pertains to a broader societal goal. The focus of governance efforts needs to be selected based on the mapping of the ecological makeup of that situation, e.g. what elements are essential for that broader societal goal. Data may be one of those elements, but there will be others that might appear just as or even more important.

## *6.3  Caveats and limitations*

All strands of the data commons literature share a number of limitations that need to be accounted for when the data commons analyses are used.

*Composition and boundaries of communities*

If the data commons are to be governed by communities, or partially composed of the communities, determining stable boundaries of these communities in the digital context will be challenging. A conceptual question to ask if what should qualify a community for being involved in the governance: the fact that it produces the data, is impacted by it, or a combination of the above. In case of the relational commons, the question would be what is the exact relationship between the data and the community, and especially how and in what way the quality of the former depends on the well-being of the latter. Which data sets and which data-related processes, 'belong' to which common and on what grounds?

---

[173] We discuss why such a transplant is not necessarily a good idea below.

The difficulty of drawing the community boundaries is further reinforced by the algorithmic context. As Purtova acknowledges, algorithmically constructed communities, being the affected stakeholders of data processing, are too dynamic and opaque to function as decision-making institutions.[174]

*Challenges of collective action*

Each strand of the data commons literature in one way or another implies collective governance. Applying collective governance to the data-related situations will face a problem of coordination of collective action. While present at the smallest scale of collective action, the problem will worsen in the context of information-related commons, e.g. digital economy, innovation or cyberspace, which are often large-scale. The larger the commons are and the bigger the group of those who potentially should be involved in its governance, the more complex the coordination will be. Data commons which exist on a local but also global scale face the challenge of collective governance on the global level. Data commons is not the only example of what Ostrom calls "global commons",[175] but these other commons like global oceans and space have so far been governed by international treaties rather than communities, and delegating commons governance to global international institutions has a significant trade-off in enforcement and effectiveness. To reflect difficult compromises, international treaties often employ very general language and - although can be binding on their state signatories - often suffer from a lack of enforcement.

---

[174] Purtova (n64) 199 et seq.
[175] E.g. Michael D. McGinnis and Elinor Ostrom, 'Design Principles for Local and Global Commons' (1992) paper presented at the Linking Local and Global Commons conference, Harvard Center for International Affairs, Cambridge, MA, 23-25 April 1992, available online at https://dlc.dlib.indiana.edu/dlc/handle/10535/5460 last accessed 26 January 2023.

*Global problems cannot be resolved by local commons.*

While resorting to large-scale commons has significant trade-offs, the success of the small-scale commons in achieving broad societal objectives, such as better control over the personal data of their members or fairer distribution of the benefits of data analysis, might be significantly limited by the fact that those local communities cannot insulate themselves from the impact of their surroundings and be "islands of decommodification"[176] or of data justice, as they are often dependent on external factors (platforms, infrastructure, laws, etc.).

*Role of state regulation*

While suggesting collective governance, all strands of the data commons literature are silent on why, if at all, collective governance is superior to public regulation. This issue is especially important when the common-pool resource is of a national, international or a global scale where the problem of coordination of collective action is salient while public governance institutions are in place. As we remember from the discussion in section 3, the collective governance of CPRs is an alternative to the market and state regulation and another tool in the governance toolbox. But when and why should we choose it, and what is the place - if any - of state regulation in the governance of common-pool resources? These questions remain unanswered, creating a risk that the role of state and state regulation in governing CPRs is minimised without a good reason.

---

[176] Evgeny Morozov. 2021. 'The Emancipatory Potential of This Discourse Rests on the Mistaken Idea That Creating Islands of Decommodification in an Ocean of Commodification — and Doing It as Efficiently as Possible — Has System-Transforming Effects. But It Doesn't. What's Missing Is the What, Not the How.' Tweet. @evgenymorozov (blog). November 17, 2021. https://twitter.com/evgenymorozov/status/1460961679383085057. Compare, P. Dardot and C. Laval, *Common: On Revolution in the 21st Century* (Bloomsbury Academic 2019) 102.

*Conceptual disarray*

The final caveat one should be aware of when applying the commons analysis to data is the conceptual disarray that dominates this field of knowledge. As highlighted above, the term "commons" is used in many meanings across different contexts and disciplines. Just in the literatures we reviewed the commons analysis is applied to data based on four distinct claims, some of which pertain to the empirical reality, while others are political. The inclusive use of the term "commons" without a doubt has value, among others, because it is conducive to creating a commons intellectual community and cross-pollination of ideas between its different parts. Some of the strands of the data commons scholarship deliver more universally applicable insights then others.[177] Yet, one should be more cautious about the precise meaning of the term in practice, e.g. when suggesting directions for policy. In particular, the IAD framework developed by Ostrom and her followers as a part of her CPR work cannot be simply adopted as a template for governance strategies when the justification for the commons analysis is not an empirical claim about the qualities of a certain data-related good, but a political claim that data ought to be held in common. The IAD framework was developed for a specific context of the common-pool resources which are complex system resources and display subtractability and low excludability. While the IAD framework does not offer a universal recipe for success for all CPRs,[178] suggesting it unconditionally for the governance of a resource that does not possess these characteristics is unfounded.

---

[177] E.g. the GKC literature delivers insights applicable to any instance of collective governance as discussed in 4.2(i).
[178] Ostrom (n49) 645.

# 7 What does it teach us about data governance?

In the following two sections, we move beyond the literature and formulate some lessons about data governance.

## 7.1 *We should adopt a dynamic characterisation of data as an economic good*

While data has conventionally been thought of as a club good and more recently a common pool resource, none of these classifications on their own accurately reflect the nature of data as a resource. The conventional neo-classical economic thinking about data as a club good adequately reflects data's characteristics (it is excludable and low in subtractability) but disregards the societal context and consequences of data production and use. The more avant-garde literature on data commons either fails to convincingly demonstrate that data is subtractable and hence is a common-pool resource (the naturalist approach) or skips the step of showing that data is a CPR altogether and proceeds immediately to applying the commons governance framework to it (the GKC). Yet, a useful account of data in terms of the commons can be constructed, if data is seen not as a CPR itself, but as a part of another larger common-pool resource ecosystem and is instrumental to its sustainability. Hess and Ostrom have constructed such an account where data is seen as a part of a larger commons of scientific knowledge, but other things can be conceptualised as this larger CPR, e.g., privacy,[179] community values, innovation, etc. The choice depends on the purpose of analysis and what - from the perspective of a given analysis - has to be preserved.

To accurately speak of data in terms of the economic goods, we argue that one does not have to pick sides (whether one sees data as a club good or a part of another larger common-pool resource) but instead should accept a dynamic classification of data as both a club good and a part of one or several other common pool resources at the same time. This classification is dynamic because one can switch between the options depending on the purposes of one's analysis. The club good characterisation and respective economic models will provide suitable guidance if the purpose of analysis is to ensure that enough data of sufficient quality is produced and available to the members of the club. The commons analytical framework is a better match where not the data but something else needs to be provided or maintained, and where data is instrumental to its sustainability.

---

[179] Henrik Skaug Sætra, 'Privacy as an Aggregate Public Good' 2020 Technology in Society 63 <https://doi.org/10.1016/j.techsoc.2020.101422> last accessed April 8 2023.

### *7.2 Governance strategies based on the neo-classical economic models of governing data as an economic good can only ensure the provision of data*

At the same time, this "data as an economic good" frame of reference adopted e.g. in the data as a club-good literature or the data as a CPR literature is not productive if one's objective is not to ensure the production or availability of *data*, but something else. Attempting to propose solutions to societal problems – even connected to data, such as fostering data economy and digital innovation – by applying the economic goods analysis to data is fated to miss the target because this type of the analysis is wired to produce solutions for the production or availability of *data* and nothing else. Therefore, if one wishes to use the economic goods frame of reference, e.g., to achieve provision of privacy, democracy or other "goods", one needs to conceptualise *those* as goods and apply the economic models to those other things. Data may or may not play a role in the governance solutions those economic models will suggest.

This is the case even for the digital economy which the European Commission has been promoting with the emerging data law. The recent regulatory avalanche has been focusing too much on data and has not taken a broader view on the data economy as a complex ecosystem resource which needs to be provided. Data might be an important but not sufficient element for the sustainability of the digital economy. In fact, because of the heavy regulation of data, the economic actors may as well reshape their digital business models so that they are not data data-dependent, and something else will become the "new oil",[180] and the data law will then miss the mark.

*7.3    Neo-classical economics engagement with data as an economic good detracts from complexity of data relations and the role data plays for collectives*

In addition to the prescriptive limitations that the economic models of data governance as a good have, utilising the economics' "goods" vocabulary in relation to data has performative effect and hence is problematic in itself. As Madison argued, conceptualisations of data are metaphorical acts with implications for the world outside.[181] As the relational commons literature suggests, this specific way of framing and treating data transforms the *relationships* human beings have with one another to goods and market objects. This changes how human beings appreciate and value these relationships. Desire or pleasure-oriented attitudes towards these newly construed goods overshadow other ways of relating and valuing,[182] and the 'market' as model and ideal becomes the benchmark with which these relationships and society more generally are evaluated.[183]

In this way, treating data as a good directly or indirectly reinforces, harmful practices of surveillance, problematic dependencies, and the exploitation of those involved in the *production* of data.[184] Furthermore, employing the neo-classical economic vocabulary may result in the market logic "transgressing" into non-market 'spheres' of life where exchange and commodification are not desirable, e.g. family and community relationships.[185]

---

[180] We owe this point to Michael Veale who commented on our research during the INFO-LEG expert workshop in Utrecht on 15 November 2022.
[181] See also Diana Coyle, *Cogs and Monsters: What Economics Is, and What It Should Be* (Princeton University Press 2021) 62-63.
[182] D. Graeber, *Toward an Anthropological Theory of Value: The False Coin of Our Own Dreams* (Palgrave 2001), 9.
[183] E.g., Marion Fourcade and Kieran Healy, 'Seeing like a Market' (2017) 15 Socio-Economic Review 9.
[184] E.g. Tamar Sharon, 'From Hostile Worlds to Multiple Spheres: Towards a Normative Pragmatics of Justice for the Googlization of Health' (2021) Medicine, Health Care and Philosophy <https://doi.org/10.1007/s11019-021-10006-7> accessed 6 July 2021; Jathan Sadowski, 'When Data Is Capital: Datafication, Accumulation, and Extraction' (2019) 6 Big Data & Society https://doi.org/10.1177/2053951718820549 last accessed April 8 2023.

### *7.4 The grass-roots collective (data) governance initiatives should not be bandaids to problems where the public regulation fails or is non-existent*

All strands of the commons literature advocate for collective governance for a variety of reasons. None of those strands is clear on if and under what circumstances the collective governance is superior and should be preferred to the state regulation. While collective governance and citizen self-organisation undeniably are of value, e.g. because they educate citizens capable of democratic participation, or because they account for the interests of communities better than a centralised state government can, we should be cautious not to overask citizens. The grass-roots citizen initiatives, e.g. data cooperatives, trusts or commons, should not be put in a position of a bandaid to fix failures of public regulation where the responsibility lies with the public authorities. The latter is clearly the case when it comes to guaranteeing fundamental rights or fixing market failures. It is interesting that a lot of the empirical studies of the commons inspired by Ostrom, but also by Ostrom herself, come from the US, a state with a particular regulatory culture where even the US Constitution is created primarily to restrain governmental action, and if the communities do not step in, nobody will. Yet, Europe is a very different context with well-developed regulation. One of the first studies by Ostrom was about how the local community governed the use of a local water basin. But in Europe environmental law is doing that.

---

[185] Coyle (n181) 34, refers to the work of Michael Sandel and Elizabeth Anderson who point at the harmful implications of the extension of the market to other spheres of life. Others, for instance, Fraser and Sassen, point at the 'expulsions' and 'cannibalistic' character of dominant economic practices. Nancy Fraser, *Cannibal Capitalism* (Verso 2022). Saskia Sassen, *Expulsions: Brutality and Complexity in the Global Economy* (The Belknap Press of Harvard university Press 2014).

This point is further reinforced by the fact that the ability of those grass-roots initiatives to make a difference is limited by the fact that those communities are embedded in larger socio-technical contexts which are often beyond their immediate control: law, technology, infrastructure, model of economy, etc.

### 7.5 *Towards a political-ecological approach to governance of digital society*

So far we have been rather critical of economic way of thinking about data as a resource. Yet, if some accounts reviewed above cannot be applied to the problems of the digital society literally, or only to a limited extent and to a limited effect, some of the insights we gained certainly provide inspiration and suggest directions and modes of governance. We find the non-naturalist commons accounts inspired by Ostrom and to some extent the relational commons literature especially compelling since they are not data-centric and account for the instrumental role of data in a digital society where values at stake are more complex than pure data provision and availability, and data is but one part of a complex ecosystem sustaining those values. Unlike Ostrom and the GKC scholars, we recognize that governance always implies normative choices that should be an integral part of the governance strategies. Below we present a short sketch of how these insights can be translated into specific lessons for the governance of the digital society, what we call "a political-ecological approach to governing the digital society".

We argued that it is not productive to focus on governance of data as an economic good if the objective is to attain a societal goal other than provision of data of sufficient quantity and quality. Therefore, if the objectives of governance are different, one needs to take a broader view on what a resource is that needs to be "provided." In many cases this resource will be complex and its provision and sustainability will depend on a number of factors, which might or might not include data. Our political-ecological approach to governance of that resource would take shape along the lines we sketch below:

1. Articulate what a value at stake is. The examples of values often cited in the data governance literature and policy are innovation, privacy, digital economy, but there could be others, like scientific or indigenous knowledge. This will be the resource or good to be governed.

2. Articulate what constitutes sustainable resource use.

3. Acknowledge that the framing of what the resource is, and hence what needs to be preserved, is political: it implies a value judgement as to what the desired resource is, to whom it should be available and under what conditions (when the resource use is considered "sustainable").[186]

4. Be responsive to discussion and contestation of this framing by those affected by it."[187]

5. Map the ecological makeup of the resource of choice with attention to any shifts in its structure due to technological changes: what are the elements essential to its sustained existence and reproduction, how do they interact, and what are the actors, technologies, values, and institutions associated with those? These elements will be potential objects of governance.

6. Devise the governance strategies recognizing these complexities and the ecological make-up of the situation.

These elements are meant to help researchers and policy makers to identify the relevant units of analysis (the 'ecosystem'), and acknowledge the moral-political nature of this action, including the ensuing duties towards those affected by the framing.

Importantly, these principles resonate with the logic of good governance (start with articulating what the governance strategy aims to achieve, map the factors that are key to the governance objective, etc.) and can guide any governance strategy, without the need to first establish that the object of governance is a common-pool resource in the sense Ostrom used it, or a commons in the GKC's sense of the word.

---

[186] For a similar account to which we partly draw here, Gijs van Maanen, From Communicating to Distributing: Studying Open Government and Open Data in the Netherlands (PhD Thesis Tilburg University 2023).

[187] On responsiveness as political norm, see Thomas Fossen, 'Constructivism and the Logic of Political Representation' (2019) 113 American Political Science Review 824.

This proposal needs to be developed further, and we aim to do this by drawing inspiration from the aforementioned work by Ostrom and Madison, Latour's work on political ecology,[188] Clarke, Friese and Washburn's work on situations, ecologies, and social worlds,[189] research on data practices,[190] and information studies research on data ecologies.[191]

# 8    Conclusion

Like mushrooms after rain, data governance models, initiatives, and proposals have been popping up in academic and policy discussions on the governance of digital society. Think for instance, of the proposals to create data trusts, data commons, or otherwise engage in data sharing. Many of these proposals, either implicitly or explicitly, understand data as an economic good. Treating data as an economic good is a particular way of relating to the digital world, which is conducive to some goals, and not or to a lesser extent to others.

This paper conducted a systematic critical review of a broad range of literature conceptualising data as an economic good and suggesting strategies based on this conceptualization. By this review we intended to achieve three things:  first, we aimed to introduce order in the chaos of the various literatures on data as an economic good, ridden with contradictions and conceptual unclarities, to make those literatures more useable both in academic and policy contexts. Second, we interrogated to what extent and for which purposes the reviewed strands of literature are helpful or unproductive to guide governance efforts. Finally, we formulated five lessons for and about data governance and governing the digital society generally.

---

[188] Bruno Latour, Politics of nature: how to bring the sciences into democracy (Harvard University Press 2004).
[189] Adele E. Clarke and Susan Leigh Star, 'The Social Worlds Framework: A Theory/Methods Package' in Edward J Hackett and others (eds), The Handbook of Science and Technology Studies (3rd edn, The MIT Press 2008).
[190] Gijs van Maanen, 'Studying Open Government Data: Acknowledging Practices and Politics' (2023) 5 Data & Policy 1.
[191] Bastiaan van Loenen, 'Towards a User-Oriented Open Data Strategy' in Bastiaan van Loenen, Glenn Vancauwenberghe and Joep Crompvoets (eds), Open Data Exposed (TMC Asser Press 2018) <https://doi.org/10.1007/978-94-6265-261-3_3> accessed 16 March 2021.

One of the main takeaways of this analysis is that the focus on data as an economic good in many governance proposals is hardwired to only result in the production of data, and not something else, even when that something is associated with data. While this paper cannot be read as supporting a broad statement that data governance is a distraction, it can be read as "the focusing on data as an economic good and using the corresponding economic models to achieve broader societal goals is a distraction". Any governance models, including data commons or cooperatives gaining in popularity, that claim to empower or help citizens to protect themselves against big tech should be re-evaluated in light of this conclusion.

Instead of making data the central focus of governance strategies of the digital society, we propose a political ecological approach towards digital governance which is defined by the ecological thinking about governance problems, i.e. it recognises the complexity of the governance goals and situations and the awareness of the political nature of framing the problems and mapping their ecological makeup. We formulated six roughly defined components that together make it an empirically and normatively adequate starting point when analysing and devising digital governance strategies. This call to approach problems in the digital society through this political-ecological lens is to be developed and applied in further research.

To conclude, governing data as an economic good has its role to play in the governance of the digital society. Provision of data of sufficient quality and quantity is a valid governance objective. Yet, governance of the digital society should not be entirely composed of the data governance measures. We should be cautious of data governance becoming a red herring which distracts from what we should care about, which hardly is the production and sharing of data alone.

**Acknowledgements**

The authors are grateful to the participants of the 'Future of regulating the digital society' workshop (15-16 November 2022), the Digital Legal Talks paper session (24 November 2022), and other discussions for helping us sharpen this paper's analysis. We are particularly grateful to Annemarie Balvert, Balázs Bodó, Charlotte Ducuing, Tommaso Fia, Laura Fichtner, Max von Grafenstein, Jörg Pohle, and Alexander Wulff.

**Acknowledgement of funding**

This contribution reports on the results of the project 'Understanding information for legal protection of people against information-induced harms' ('INFO-LEG'). This project has received funding from the European Research Council (ERC) under the European Union's Horizon 2020 research and innovation programme (grant agreement No 716971). The paper reflects only the authors' views and the ERC is not responsible for any use that may be made of the information it contains. The funding source had no involvement in study design, in the collection, analysis and interpretation of data, in the writing of the report, and in the decision to submit the article for publication.